\documentclass[%
 reprint,
 superscriptaddress,
 amsmath,amssymb,
 aps,
 prl,
]{revtex4-2}
\usepackage{graphicx}
\usepackage{dcolumn}
\usepackage{bm}
\usepackage{siunitx}
\usepackage{lineno}
\usepackage{xcolor}
\usepackage{comment}
\usepackage{adjustbox}
\usepackage[utf8x]{inputenc}
\usepackage{ucs}
\usepackage{lineno}
\usepackage[normalem]{ulem}
\usepackage{subfigure}

\begin{document}
\title{Search for Light Dark Matter-Electron Scatterings in the PandaX-II Experiment}

\def\shKeyLab{School of Physics and Astronomy, Shanghai Jiao Tong University, Key Laboratory for Particle Astrophysics and Cosmology (MOE), Shanghai Key Laboratory for Particle Physics and Cosmology, Shanghai 200240, China}
\def\BUAA{School of Physics, Beihang University, Beijing 100191, China}
\def\USTClab{State Key Laboratory of Particle Detection and Electronics, University of Science and Technology of China, Hefei 230026, China}
\def\USTCdep{Department of Modern Physics, University of Science and Technology of China, Hefei 230026, China}
\def\BUAALab{International Research Center for Nuclei and Particles in the Cosmos \& Beijing Key Laboratory of Advanced Nuclear Materials and Physics, Beihang University, Beijing 100191, China}
\def\pku{School of Physics, Peking University, Beijing 100871, China}
\def\YaLongSD{Yalong River Hydropower Development Company, Ltd., 288 Shuanglin Road, Chengdu 610051, China}
\def\IAP{Shanghai Institute of Applied Physics, Chinese Academy of Sciences, 201800 Shanghai, China}
\def\CHEPpku{Center for High Energy Physics, Peking University, Beijing 100871, China}
\def\SDUdep{Research Center for Particle Science and Technology, Institute of Frontier and Interdisciplinary Science, Shandong University, Qingdao 266237, Shandong, China}
\def\SDUlab{Key Laboratory of Particle Physics and Particle Irradiation of Ministry of Education, Shandong University, Qingdao 266237, Shandong, China}
\def\UMD{Department of Physics, University of Maryland, College Park, Maryland 20742, USA}
\def\TDLee{Tsung-Dao Lee Institute, Shanghai Jiao Tong University, Shanghai, 200240, China}
\def\MESJTU{School of Mechanical Engineering, Shanghai Jiao Tong University, Shanghai 200240, China}
\def\SYU{School of Physics, Sun Yat-Sen University, Guangzhou 510275, China}
\def\NKU{School of Physics, Nankai University, Tianjin 300071, China}
\def\FDU{Key Laboratory of Nuclear Physics and Ion-beam Application (MOE), Institute of Modern Physics, Fudan University, Shanghai 200433, China}
\def\USST{School of Medical Instrument and Food Engineering, University of Shanghai for Science and Technology, Shanghai 200093, China}
\def\SJTUSC{Shanghai Jiao Tong University Sichuan Research Institute, Chengdu 610213, China}
\def\Princeton{Physics Department, Princeton University, Princeton, NJ 08544, USA}
\def\MIT{Department of Physics, Massachusetts Institute of Technology, Cambridge, MA 02139, USA}
\def\SARI{Shanghai Advanced Research Institute, Chinese Academy of Sciences, Shanghai 201210, China}

\affiliation{\shKeyLab}

\author{Chen Cheng}\affiliation{\SYU}
\author{Pengwei Xie}\affiliation{\TDLee}
\author{Abdusalam Abdukerim}
\author{Wei Chen}\affiliation{\shKeyLab}
\author{Xun Chen}\affiliation{\shKeyLab}\affiliation{\SJTUSC}
\author{Yunhua Chen}\affiliation{\YaLongSD}
\author{Xiangyi Cui}\affiliation{\TDLee}
\author{Yingjie Fan}\affiliation{\NKU}
\author{Deqing Fang}
\author{Changbo Fu}\affiliation{\FDU}
\author{Mengting Fu}\affiliation{\pku}
\author{Lisheng Geng}\affiliation{\BUAA}\affiliation{\BUAALab}
\author{Karl Giboni}
\author{Linhui Gu}\affiliation{\shKeyLab}
\author{Xuyuan Guo}\affiliation{\YaLongSD}
\author{Ke Han}\affiliation{\shKeyLab}
\author{Changda He}\affiliation{\shKeyLab}
\author{Shengming He}\affiliation{\YaLongSD}
\author{Di Huang}\affiliation{\shKeyLab}
\author{Yan Huang}\affiliation{\YaLongSD}
\author{Yanlin Huang}\affiliation{\USST}
\author{Zhou Huang}\affiliation{\shKeyLab}
\author{Xiangdong Ji}\affiliation{\UMD}
\author{Yonglin Ju}\affiliation{\MESJTU}
\author{Shuaijie Li}\affiliation{\TDLee}
\author{Qing  Lin}\affiliation{\USTClab}\affiliation{\USTCdep}
\author{Huaxuan Liu}\affiliation{\MESJTU}
\author{Jianglai Liu}\email[Spokesperson: ]{jianglai.liu@sjtu.edu.cn}\affiliation{\shKeyLab}\affiliation{\TDLee}\affiliation{\SJTUSC}
\author{Liqiang Liu}\affiliation{\YaLongSD}
\author{Xiaoying Lu}\affiliation{\SDUdep}\affiliation{\SDUlab}
\author{Wenbo Ma}\affiliation{\shKeyLab}
\author{Yugang Ma}\affiliation{\FDU}
\author{Yajun Mao}\affiliation{\pku}
\author{Yue Meng}\email[Corresponding author: ]{mengyue@sjtu.edu.cn}\affiliation{\shKeyLab}\affiliation{\SJTUSC}
\author{Parinya Namwongsa}\affiliation{\shKeyLab}
\author{Kaixiang Ni}\affiliation{\shKeyLab}
\author{Jinhua Ning}\affiliation{\YaLongSD}
\author{Xuyang Ning}\affiliation{\shKeyLab}
\author{Xiangxiang Ren}\affiliation{\SDUdep}\affiliation{\SDUlab}
\author{Nasir Shaheed}\affiliation{\SDUdep}\affiliation{\SDUlab}
\author{Changsong Shang}\affiliation{\YaLongSD}
\author{Guofang Shen}\affiliation{\BUAA}
\author{Lin Si}\affiliation{\shKeyLab}
\author{Andi Tan}
\affiliation{\UMD}
\author{Anqing Wang}\affiliation{\SDUdep}\affiliation{\SDUlab}
\author{Hongwei Wang}\affiliation{\SARI}
\author{Meng Wang}\affiliation{\SDUdep}\affiliation{\SDUlab}
\author{Qiuhong Wang}\affiliation{\FDU}
\author{Siguang Wang}\affiliation{\pku}
\author{Wei Wang}\affiliation{\SYU}
\author{Xiuli Wang}\affiliation{\MESJTU}
\author{Zhou Wang}\affiliation{\shKeyLab}\affiliation{\SJTUSC}
\author{Mengmeng Wu}\affiliation{\SYU}
\author{Shiyong Wu}\affiliation{\YaLongSD}
\author{Weihao Wu}
\author{Jingkai Xia}\affiliation{\shKeyLab}
\author{Mengjiao Xiao}
\affiliation{\UMD}
\author{Xiang Xiao}\affiliation{\SYU}
\author{Binbin Yan}\affiliation{\shKeyLab}
\author{Jijun Yang}
\author{Yong Yang}\affiliation{\shKeyLab}
\author{Chunxu Yu}\affiliation{\NKU}
\author{Jumin Yuan}\affiliation{\SDUdep}\affiliation{\SDUlab}
\author{Ying Yuan}\affiliation{\shKeyLab}
\author{Xinning Zeng}\affiliation{\shKeyLab}   
\author{Dan Zhang}\affiliation{\UMD}
\author{Tao Zhang}
\author{Li Zhao}\affiliation{\shKeyLab}
\author{Qibin Zheng}\affiliation{\USST}
\author{Jifang Zhou}\affiliation{\YaLongSD}
\author{Ning Zhou}\affiliation{\shKeyLab}
\author{Xiaopeng Zhou}\affiliation{\BUAA}

\collaboration{PandaX-II Collaboration}
\noaffiliation

\date{\today}

\begin{abstract}
We report constraints on light dark matter through its interactions with shell electrons in the PandaX-II liquid xenon detector 
with a total 46.9 tonne$\cdot$day exposure. To effectively search for these very low energy electron recoils, 
ionization-only signals are selected from the data. 
1821 candidates are identified within ionization signal range between 
50 to 75 photoelectrons, corresponding to a mean electronic recoil energy from 0.08 to 0.15~keV.
The 90\% C.L. exclusion limit on the scattering cross section 
between the dark matter and electron is calculated with systematic uncertainties properly taken into account. Under the assumption of point interaction, we provide the world's most stringent limit 
within the dark matter mass range from 
15 to 30~$\rm{MeV/c^2}$, with the corresponding cross section from $2.5\times10^{-37}$ to $3.1\times10^{-38}$ cm$^2$.

\end{abstract}

\maketitle
\begin{figure*}
    \includegraphics[width=12cm]{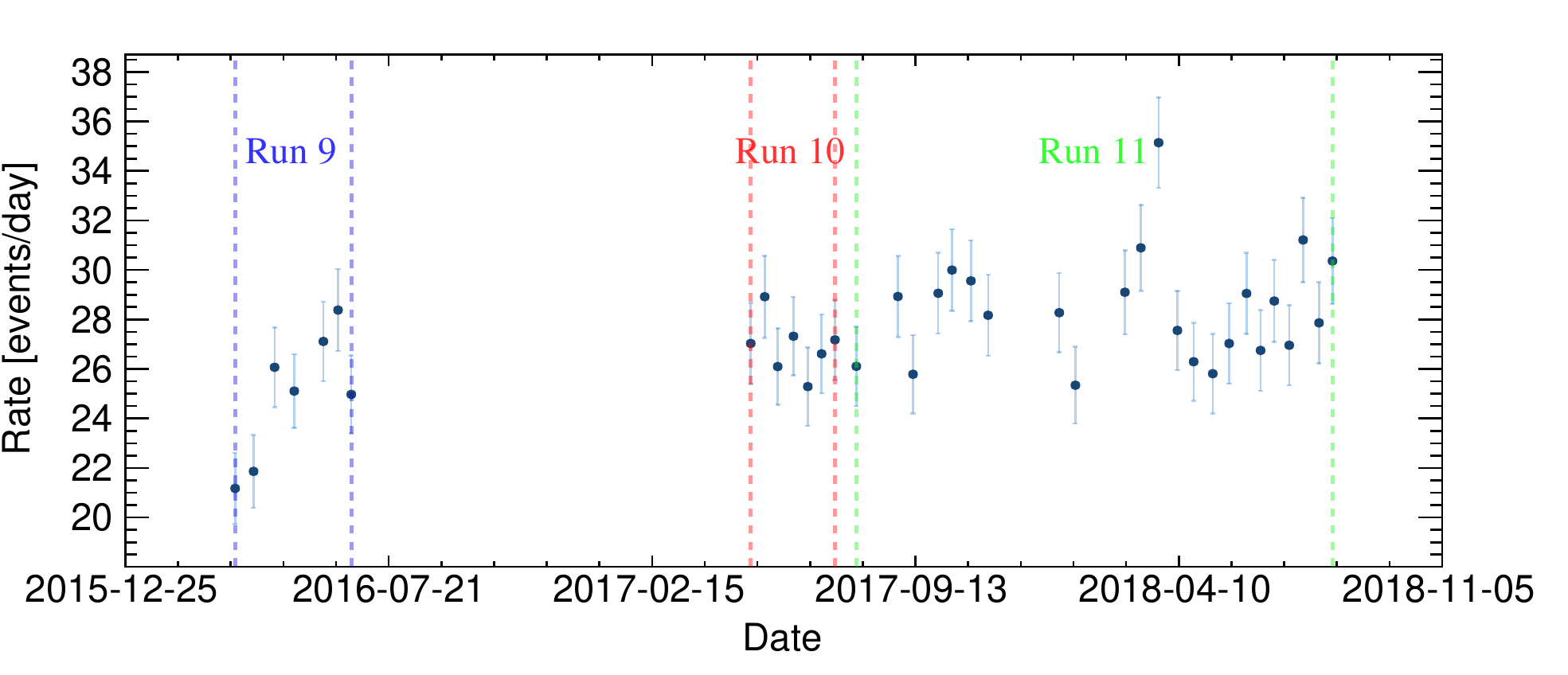}
    \caption{\label{fig:timeevolution}Event rate of the US2 signals with charge range from 50 to 200 PE 
    }
\end{figure*}%
The existence of dark matter (DM) is established by overwhelming evidence in cosmological and astronomical observations~\cite{BERTONE2005279review}. 
Possible interactions between DM particles and baryonic matter have been searched for in underground laboratories using ultra-low background detectors by directly 
detecting recoil signals~\cite{Liu:2017drfPandaXII,Undagoitia_2015_DMreview}.
DM within a mass range between $\rm{GeV/c^2}$ to $\rm{TeV/c^2}$ have been mostly searched for via its elastic scattering off atomic nucleus~\cite{Tan:2016dizPandaXII, 
Aprile:2018dbl_xenon1tNR, Akerib:2016vxi_LUXNR, Agnes:2018fwg_darksideNR,Ajaj:2019imk,Agnese:2013jaa,Jiang:2018pic,Abdelhameed:2019hmk,Amole:2017dex}.
The scatterings of DM within this mass range with electrons, while still possible, is difficult to be  kinematically probed as the energy of electron recoils (ERs) is 
suppressed by the smallness of electron mass. 
For sub-GeV light DM, on the other hand, 
the nuclear recoil energy becomes much more difficult to detect with conventional detection techniques. 
It was
realized that these light DM can scatter with shell electrons, which may subsequently produce sufficiently large ionization signals in the detector~\cite{PhysRevD.85.076007DMmodel}. Such DM-electron scatterings open up a new experimental paradigm, which has since been pursued by numbers of groups~\cite{PhysRevLett.109.021301, PhysRevLett.123.251801, PhysRevLett.121.111303darkside, PhysRevLett.123.181802, Barak:2020fql, Emken_2019}.

The PandaX-II experiment~\cite{Tan:2016dizPandaXII,Fu:2016ega,Fu:2017lfc,Chen:2017cqcPandaXII,Cui:2017nnnPandaXII,Xia:2018qgsPandaXII,Ren:2018gy,Wang:2019opt,Ni:2019kms,finalcpc,axion}, located in the China Jinping Underground Laboratory (CJPL), utilizes a dual-phase Time Projection Chamber (TPC), which contains an active 
cylindrical target with 580~$\rm{kg}$ of liquid xenon (LXe).
DM scattering produces prompt scintillation photons ($S1$) in the liquid region, and ionized electron signal ($S2$) 
through delayed electroluminescence in the gaseous region. 
Both $S1$ and $S2$ can be detected by two arrays of Hamamatsu R11410-20 photomultiplier tubes (PMTs) located at the top and bottom of the cylindrical volume, with a corresponding photon detection efficiency and electron extraction efficiency around 10\% and 50\%, respectively~\cite{finalcpc}. 
Conventional analysis searched for 
recoil energy from keV and above by requiring correlated $S1$
and $S2$ signals. In this work, we search for 
light DM through its scattering off shell electrons, which would produce 
energy deposition in the order of 100~eV~\cite{PhysRevD.85.076007DMmodel}. In this region, 
$S1$ would be nearly invisible, but unpaired $S2$ (US2) signals can still be used to probe
these scatterings effectively.
\begin{figure}[!bh]
    \includegraphics[width=8.6cm]{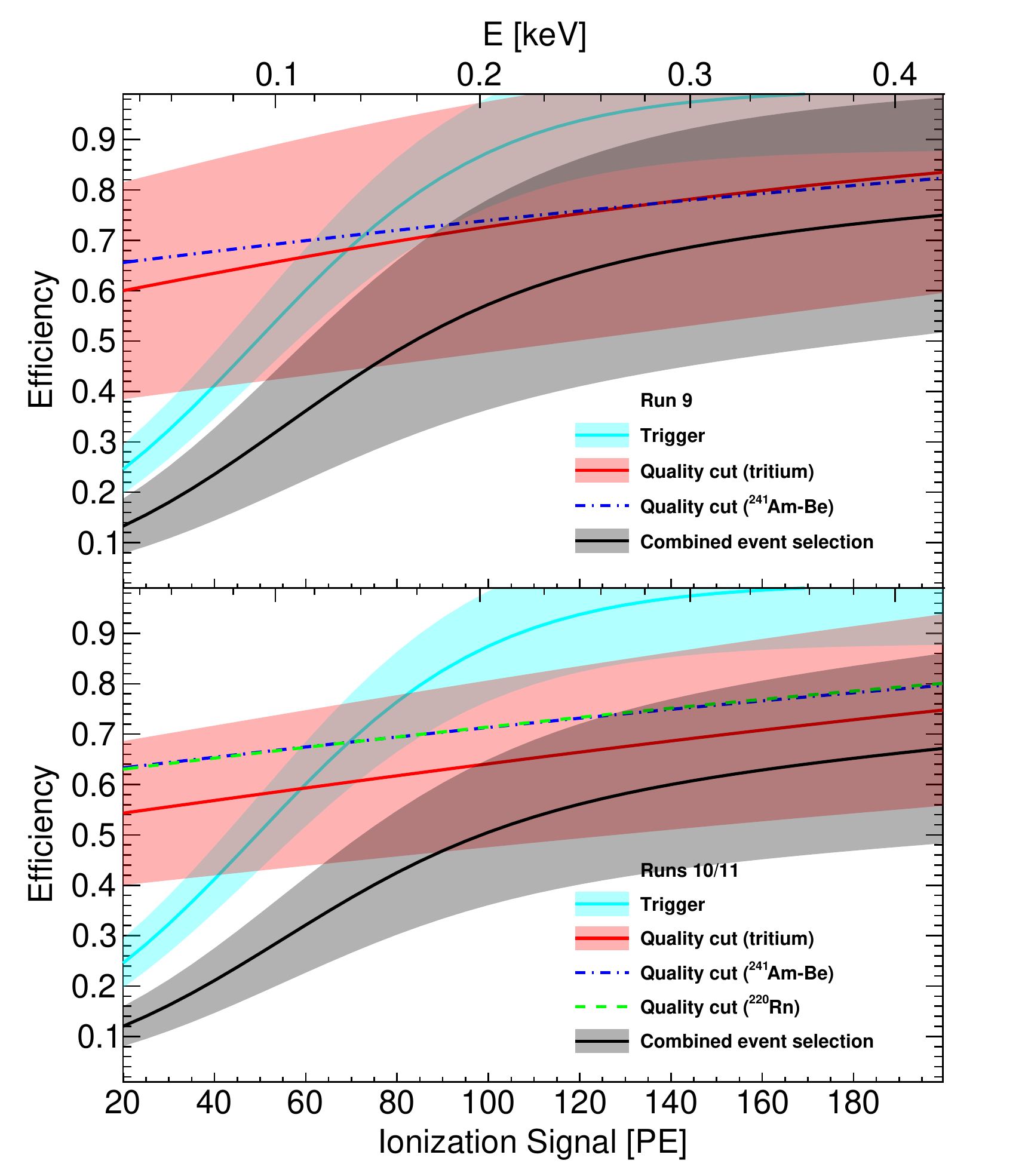}
    \caption{Data quality cut efficiency (red), trigger efficiency (blue) and combined efficiency (black) vs. detected ionization signals in the Run 9 (top) and Runs 10/11 (bottom). Details about the uncertainties are described in \cite{supplementary}. The quality cut efficiencies (dashed) from $^{241}$Am-Be and $^{220}$Rn calibration runs are used to validate that from tritium. The upper axes
    correspond to mean ER energy computed using the constant model ($f_e=0.83$ and $E = 13.7~{\rm eV}\times S2/{\rm SEG/EEE}/f_e$, see Fig.~\ref{fig:convertion}).}
    \label{fig:eff}
\end{figure}%

All DM search runs (Runs 9, 10, and 11), and calibration runs with tritium, $^{220}$Rn, and $^{241}$Am-Be sources are used in this analysis. 
Most $S2$ related data quality cuts are inherited from previous analysis~\cite{finalcpc}, including a $S2$ isolation time cut of 0.02 s~\cite{supplementary}, 
a position reconstruction quality cut to suppress single electron pileup due to multiple scatterings, and a gate event cut which searches sub-threshold charge between 1.5-3.5  $\rm{\mu s}$ window preceding an US2. Three cuts are tightened, including the full-width-10\%-maxima, the rising edge (defined as the ratio of the charge in the first 1~${\rm{\mu}}$s to the total), and the top/bottom charge ratio of the US2s, based on the distribution of the ER calibration data~\cite{supplementary}.
Although the US2 data are not intentionally blinded, which allows a confirmation of the validity of the basic data quality cuts from previous analysis, cuts that are most relevant to the selection efficiency (S2 width, rising edge, and top/bottom ratio) are developed entirely based on multiple sets of calibration data. Events are further selected with a more conservative radial cut so events within 15 cm from the wall are removed to avoid un-paired field cage surface events, leading to a 117.0 $\pm$ 6.6~kg fiducial mass of LXe. 
The uncertainty is dominated by the 4.2 mm radial position resolution, estimated using different reconstruction algorithms.
The event distribution within the fiducial volume (FV) is mostly uniform~\cite{supplementary}. The radial cut efficiency, since it has little dependence on the charge and characters of the US2s, is factored in the FV and exposure.
The time evolution of the surviving candidates in the three 
DM search runs with charge range from 50 to 200 photoelectron (PE) is shown in Fig.~\ref{fig:timeevolution}, and appears to be reasonably stable within a total time span of two years.

The combined event selection efficiency for US2 events 
is a product of the trigger efficiency 
and data quality cut efficiency. 
PandaX-II utilizes an FPGA-based realtime trigger system,
with its efficiency directly measured
from events with multiple $S2$s~\cite{Wu:2017cjlPandaXII}. The uncertainty of the trigger efficiency is estimated 
to be 12.0\% (fractional unless otherwise specified) by making comparison of the efficiencies for S2s with different widths.
The data quality cut efficiency is obtained from the tritium calibration run, given that the distributions of the low energy electron recoils are closer to the desired DM-electron scattering events.
The overall efficiency is a product of each individual cut efficiency, estimated by the ratio of survived events with all-cuts-applied to those with all-but-this-cut, also known as 
the ``N/(N-1)" approach. An alternative approach (``N/all"), which takes the ratio of all-cuts-applied to those with only the inherited basic data quality cuts, yields good agreement, indicating that the correlation between cuts can be ignored.
For comparison, the data quality efficiency curves obtained using $^{241}$Am-Be (Runs 9/10/11) and $^{220}$Rn (Runs 10/11) data are overlaid. The residual systematic uncertainty of the data quality cuts is estimated to be 13.6\%, consisting of the effect of non-source contamination in the calibration (10.0\%), the dependence of selection efficiency to event energy and position distributions thereby the shapes of US2s (9.2\%), and the S2 identification efficiency (1.5\%)~\cite{supplementary}. The trigger, data quality, and the combined efficiencies are shown 
in Fig.~\ref{fig:eff} for US2 events with charge range between 20 to 200 PE. 
\begin{figure}[!h]
    \includegraphics[width=8.6cm]{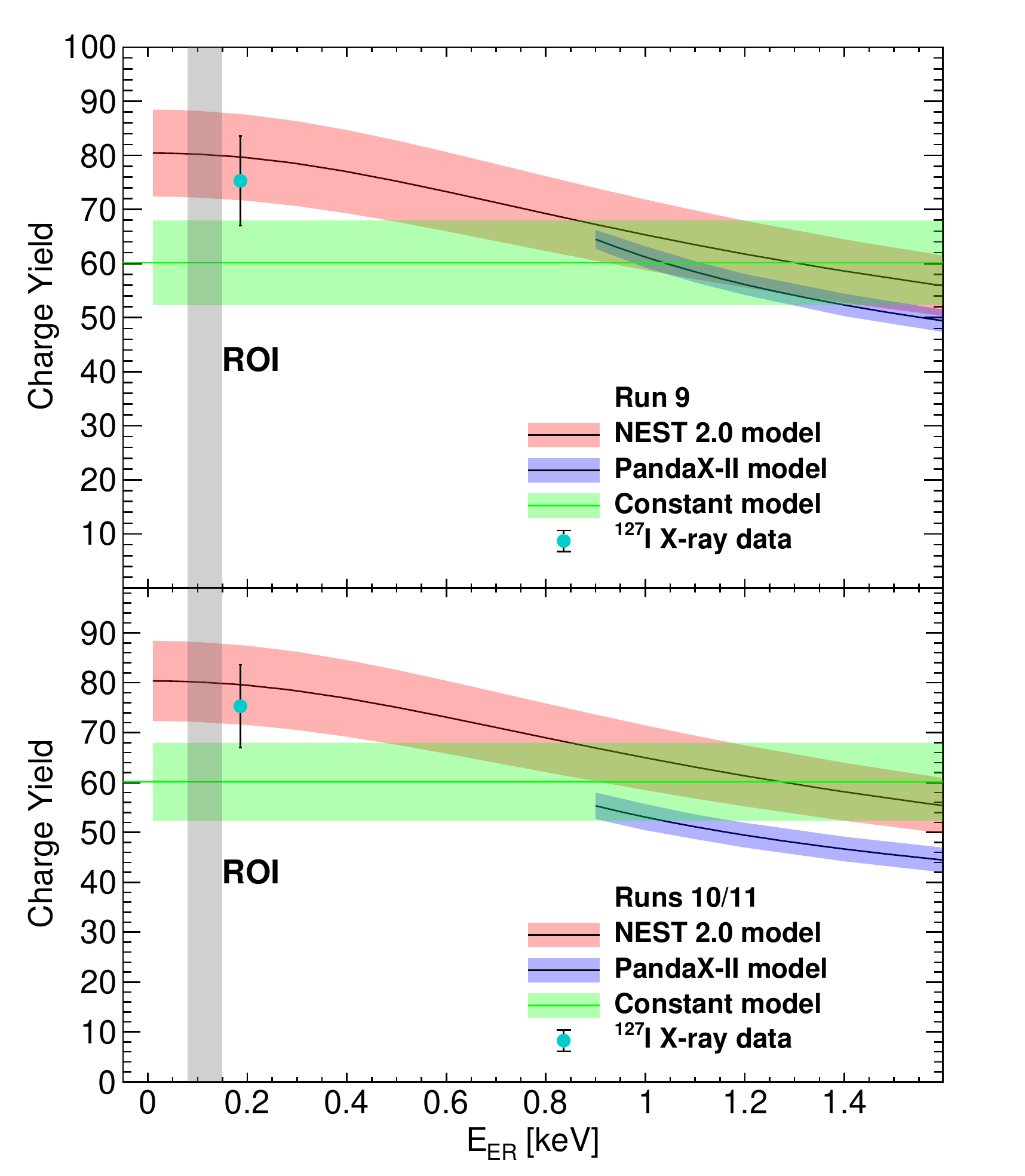}
    \caption{ Charge yield vs. electron recoil energy for three DM search runs for NEST 2.0~\cite{PhysRevD.103.012002,NEST2calculator}, the constant model~\cite{PhysRevB.76.014115}, and the PandaX-II  model~\cite{binbin}, with uncertainty
    bands obtained from original publications.
    The lowest energy measurement at 186~eV~\cite{LXeERCY:PhysRevD.96.112011} is also shown. The shaded vertical band indicates the mean energy range corresponding to the ROI (50-75~PE).
    }
    \label{fig:convertion}
\end{figure}%

As mentioned in the introduction, the light DM-electron 
scatterings produce sub-keV recoil energy, therefore 
knowledge of the photon and ionization productions in LXe in this energy regime is required. Three 
independent signal response models are compared, all under the 
standard $W$ value of 13.7~eV to produce either 
a photon or electron~\cite{Lenardo_2015}: 1) the Noble Element Simulation Technique (NEST 2.0) model~\cite{NEST2calculator, PhysRevD.103.012002}, 2) the constant model in which the fraction of ionization to total quanta is $f_{e}=0.83$~\cite{PhysRevB.76.014115} with no energy dependence and without recombination effects, and 3) the PandaX-II model, obtained by fitting the tritium calibration data with correlated $S1$ and $S2$ but with lower energy truncated to 0.9~keV due to threshold~\cite{binbin}.
The charge yield vs. recoil
energy is shown in Fig.~\ref{fig:convertion} for 
Run 9 and Runs 10/11 under a drift field of 400 and 318~V/cm, respectively. 
In general, the constant model predicts smaller charge yield in comparison with NEST 2.0. 
For Run 9, the PandaX-II model agrees with other two models within 1$\sigma$ at 0.9~keV. 
On the other hand, for Runs 10/11, the PandaX-II model agrees with the constant model, but has slight tension with NEST 2.0.
Therefore, the constant model is selected as the nominal model in this analysis to conservatively estimate the number of primary ionized electrons, as well as to be consistent with other analysis. But one should keep in mind that for the charge yield in a liquid xenon detector, the lowest energy measurement is only recently made at 186~eV at 180~V/cm~\cite{LXeERCY:PhysRevD.96.112011}. The spectrum of detected ionization signals, i.e. US2 events in PE, can then be predicted 
based on the measured detector parameters~\cite{finalcpc}, listed in Table~\ref{tab:pars} for convenience, and the efficiencies in Fig.~\ref{fig:eff}. The electron lifetime, i.e. the attenuation of ionized electrons due to electronegative impurities, is incorporated in the DM-electron signal model, by randomly distributing vertices in the 60 cm drift length, and then by applying the measured variations of the electron lifetime throughout the data taking.

\begin{table}[!th]
\caption{The PandaX-II detector parameters, including
electron extraction efficiency (EEE), single electron gain (SEG) and its measured resolution ($\sigma_{\rm SE}$)~\cite{finalcpc}. They are used to estimate the relation between ER energy and detected ionization electrons. }
\label{tab:pars}
\begin{ruledtabular}
\begin{tabular}{cccc} 
 & Run 9 & Run 10 & Run 11\\ \hline
EEE (\%) & $46.4\pm1.4$& $50.8\pm2.1$ & $47.5\pm2.0$\\ \hline
SEG (PE) & $24.4\pm0.4$ & $23.7\pm0.8$ & $23.5\pm0.8$\\ \hline
$\sigma_{\rm SE}$ (PE)  & 8.3 & 7.8 & 8.1 \\
\end{tabular}
\end{ruledtabular}
\end{table}
\begin{table*}[!th]
\caption{\label{tab:candidates}
The number of US2 candidates, exposure, and known ER background events for the three DM search runs. The span 1 and span 2 of the Run 11 are listed separately due to the different background rates. ROI is chosen as from 50 PE and 75 PE, corresponding to a mean ER energy between 0.08 to 0.15~keV. 
The flat ER background includes $^{85}$Kr, $^{222}$Rn, $^{220}$Rn, material ER, solar neutrino and $^{136}$Xe~\cite{finalcpc}.}
 \scalebox{1}{
\begin{ruledtabular}
\begin{tabular}{cccccc} 
& Run 9 & Run 10 & Run 11 span 1  & Run 11 span 2  & Total \\ \hline
Exposure $[\rm tonne \cdot \rm day]$ & 9.3 & 9.0 &  \multicolumn{2}{c}{28.6} & 46.9\\ \hline
DM-electron candidates [events]  & 287 & 340 &  \multicolumn{2}{c}{1194} & 1821 \\ \hline
Flat ER background [events]  & 0.8 & 0.2 & 0.3 & 0.6 & 1.8\\ \hline
Tritium background [events] & 0 & 0.1 & 0.2 & 0.3 & 0.6 \\
\end{tabular}
\end{ruledtabular}
 }
\end{table*}%

\begin{figure}[!th]
\includegraphics[width=8.6cm]{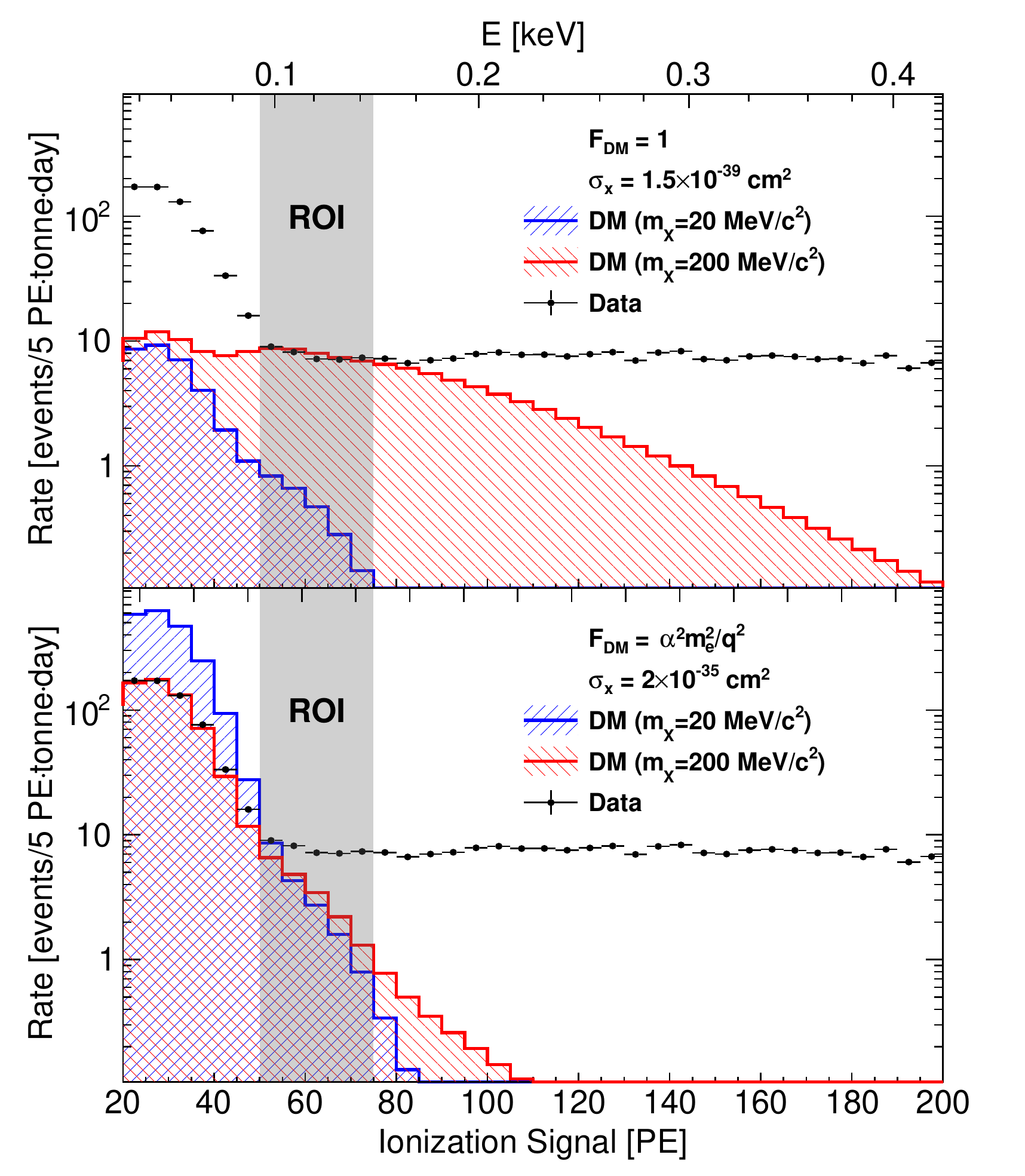}
\caption{\label{fig:DM_electron_spectrum} Detected ionization signals (US2, black histograms), and expected
signals from 
DM-electron scatterings with $F_{\rm DM}$=1 (upper) and 
$\alpha^2 m^2_e/q^2$ (lower), with 
blue (red) histogram corresponding to a DM mass of 20 $\rm{MeV/c^2}$ 200 $\rm{MeV/c^2}$). The gray shadow shows the ROI of this analysis. The excess in the data peaking at $\sim$25~PE are single electron events, likely due to stray electrons in LXe.}
\end{figure}%
\begin{figure}[!hb]
  \includegraphics[width=\linewidth]{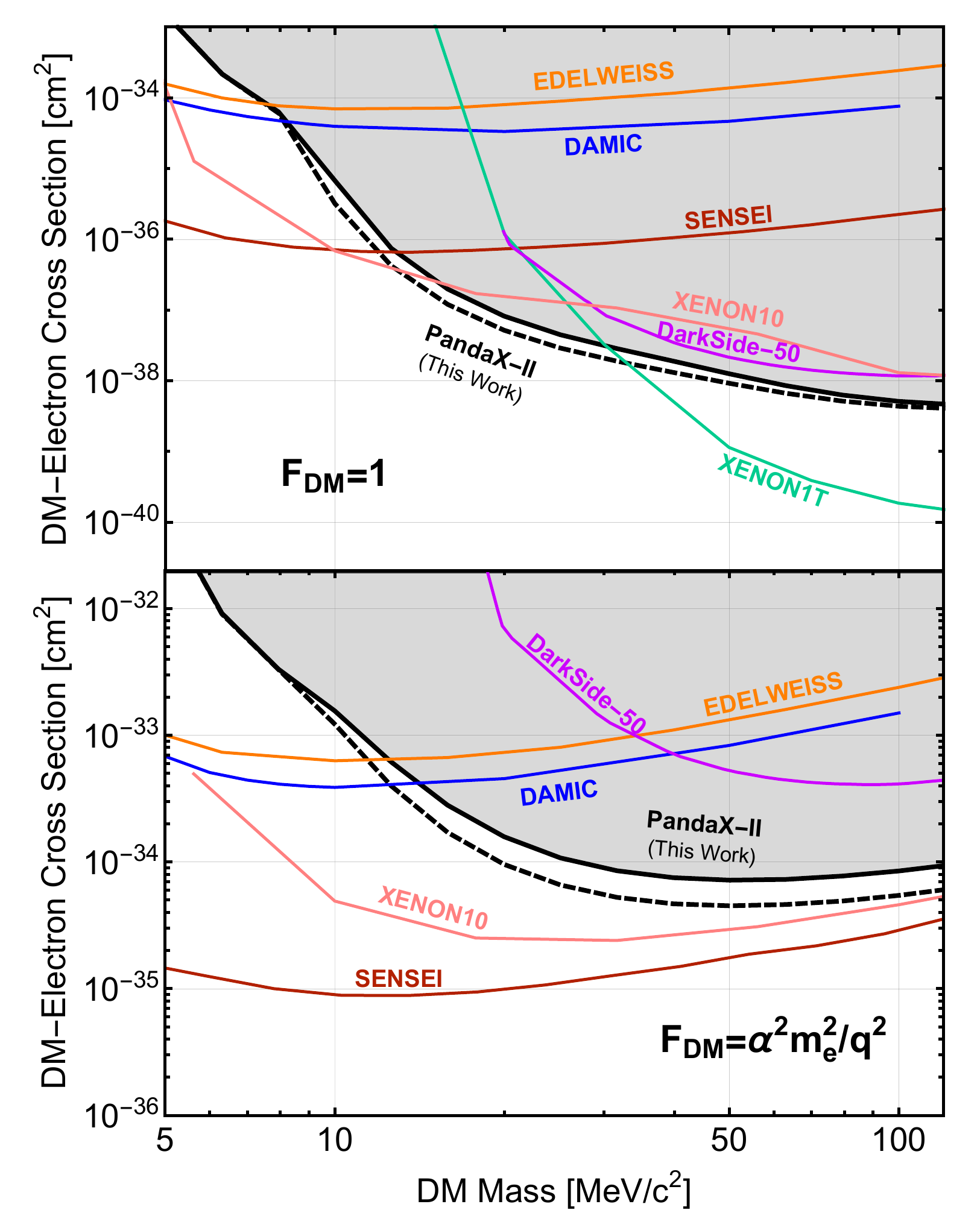}
\caption{$90\%$ C.L. upper limits (solid: the constant model, dashed: NEST 2.0) on DM-electron scattering cross section from PandaX-II data for $F_{\rm DM}=1$ (upper) and $F_{\rm DM}=\alpha^2m^2_e/q^2$ (lower). Results from XENON1T~\cite{PhysRevLett.123.251801}, XENON10~\cite{PhysRevD.96.043017DMmodel}, DarkSide-50~\cite{PhysRevLett.121.111303darkside}, SENSEI~\cite{Barak:2020fql}, DAMIC~\cite{PhysRevLett.123.181802}, EDELWEISS~\cite{Arnaud:2020svb} are also shown for comparison.}
\label{fig:comp} 
\end{figure}%
The predicted rates and recoil energy spectra of DM-electron scatterings in LXe target are 
calculated following the procedures in Refs.~\cite{calculator, PhysRevD.85.076007DMmodel}. It is recently pointed out that such calculation contains non-negligible theoretical uncertainties due to relativistic effect for inner-shell ionization and final-state free electron wave function~\cite{RelativisticEffect:PhysRevD.102.123025}. We adopt the procedures of Refs.~\cite{PhysRevD.85.076007DMmodel, calculator} nevertheless, so that our results can be directly compared to previous work.
Electron shell corrections are applied in the signal model and the minimum of the additional quanta range is taken (e.g. three additional quanta for $4s^2$ shell from Table II of Ref.~\cite{PhysRevD.96.043017DMmodel}) to conservatively estimate the ionization.
Two benchmark interaction models, the point-like interaction with form-factor $F_{\rm DM}=1$, and light mediator with $F_{\rm DM}=\alpha^2m^2_e/q^2$, are considered, with their corresponding ionization cross sections from xenon atoms computed.
The US2 candidates from the data are overlaid with 
the predicted ionization spectra for the two models 
with DM masses of 20 and 200 ${\rm MeV/c^{2}}$, respectively, in Fig.~\ref{fig:DM_electron_spectrum}.
The region-of-interest (ROI) is chosen to be 50 to 75~PE. 
The lower cut is set to keep at least 50\% trigger efficiency, and the higher cut is set to enclose the most high energy tail for the DM benchmark DM mass of 20~$\rm MeV/c^2$.
The number of candidates are summarized in Table~\ref{tab:candidates} for the three DM search runs, 
which are significantly higher than the dominating known ER background with approximately flat spectrum at low energy (flat ER) and 
 the tritium contribution~\cite{finalcpc}. 
We note that in comparison to a standard $S1$-$S2$ analysis, the US2 analysis reported here reduces the energy threshold significantly, but is more vulnerable to un-modeled background contamination due to the absence of $S1$. To be conservative, we
assume that all detected candidates are DM-electron 
scattering events in the ROI. The 90\% C.L. upper limit is derived using a cut-and-count approach via toy Monte Carlo, incorporating the statistical uncertainty and the systematic uncertainties due to the efficiency (Fig.~\ref{fig:eff}), detector responses (Table~\ref{tab:pars}), and the FV. The exclusion curves of DM-electron scattering cross section under the two benchmark interaction models are shown in 
Fig.~\ref{fig:comp}, 
assuming both the constant and NEST 2.0 signal response models. For comparison, results from earlier experiments are also overlaid in Fig.~\ref{fig:comp}. Due to the achieved 50 PE analysis threshold ($\sim$0.08~keV), 
the most stringent exclusion limit for DM-electron interaction is given for the point-interaction 
within the DM mass range from 
15 to 30~$\rm{MeV/c^2}$, with the corresponding cross section from $2.5\times10^{-37}$ to $3.1\times10^{-38}$ cm$^2$. 
At 25~$\rm{MeV/c^2}$, our result is a few times more constraining than that from XENON10 and XENON1T which used the same xenon target and the ionization-only channel~\cite{PhysRevLett.123.251801, PhysRevD.96.043017DMmodel}. 
Alternative choice of the NEST 2.0 model would increase the ionization yield relative to the constant model at a given energy, leading to a more constraining limit. 
In the near future, the PandaX-4T experiment will be under operation. With larger exposure, lower background and a lower threshold using triggerless readout \cite{hongguang}, PandaX-4T will provide more sensitive search for DM-electron scatterings. \\
\indent
This project is supported in part by Office of Science and
Technology, Shanghai Municipal Government (grant No. 18JC1410200), 
a grant from the Ministry of Science and Technology of
China (No. 2016YFA0400301), grants from National Science
Foundation of China (Nos. 12005131, 11905128, 12090061, 11775141), and a grant from Sichuan Science and Technology Program (No.2020YFSY0057). We thank supports from Double First Class Plan of
the Shanghai Jiao Tong University. We also thank the sponsorship from the
Chinese Academy of Sciences Center for Excellence in Particle
Physics (CCEPP), Hongwen Foundation in Hong Kong, and Tencent
Foundation in China. Finally, we thank the CJPL administration and
the Yalong River Hydropower Development Company Ltd. for
indispensable logistical support and other help. We thank Rouven Essig for the helpful email discussions about the theoretical spectrum at the beginning of this project. We also thank Matthew Szydagis for the helpful email communication about the NEST 2.0 model.

\bibliographystyle{apsrev4-2.bst}
\bibliography{apssamp}

\end{document}